\documentclass{article}
\usepackage{spconf,amsmath,graphicx}
\usepackage{multirow}

\usepackage{babel}
\usepackage[table]{xcolor}
\usepackage{collcell}
\usepackage{hhline}
\usepackage{pgf}
\usepackage{url}

\title{Multi-Slice Net: A novel light weight framework for COVID-19 Diagnosis}
\name{Harshala Gammulle, Tharindu Fernando, Sridha Sridharan, Simon Denman, Clinton Fookes\vspace{-2mm}}

\address{The Signal Processing, Artificial Intelligence and Vision Technologies (SAIVT), \\ Queensland University of Technology, Australia.}

\begin{document}
%
\maketitle
\begin{abstract}
This paper presents a novel lightweight COVID-19 diagnosis framework using CT scans. Our system utilises a novel two-stage approach to generate robust and efficient diagnoses across heterogeneous patient level inputs. We use a powerful backbone network as a feature extractor to capture discrimin-ative slice-level features. These features are aggregated by a lightweight network to obtain a patient level diagnosis. The aggregation network is carefully designed to have a small number of trainable parameters while also possessing sufficient capacity to generalise to diverse variations within different CT volumes and to adapt to noise introduced during the data acquisition. We achieve a significant performance increase over the baselines when benchmarked on the SPGC COVID-19 Radiomics Dataset, despite having only 2.5 million trainable parameters and requiring only 0.623 seconds on average to process a single patient's CT volume using an Nvidia-GeForce RTX 2080 GPU. 
\end{abstract}
\section{Introduction}
\label{sec:intro}

Although Reverse Transcription Polymerase Chain Reaction (RT-PCR) is considered the global standard SARS-CoV-2 (COVID-19) diagnosis, this test is very time consuming and has a high false negative rate, which in turn yields significant challenges in preventing the spread of the infection \cite{heidarian2020ct, zheng2020deep}. As such, Computed Tomography (CT) imaging has been identified as a fast, simple and reliable diagnosis tool due to the existence of discriminative patterns associated with the COVID-19 infection within the CT scans. However, recent literature has shown that COVID-19 lung manifestations show substantial similarities with Community Acquired Pneumonia (CAP), complicating the  diagnosis process \cite{heidarian2020ct}. 

To this end several deep learning based frameworks have been introduced to automate diagnosis, where models are trained to uncover discriminative patterns embedded within the data and which cannot be identified by the naked-eye. This paper presents the QUT SAIVT team's \footnote{\url{https://research.qut.edu.au/saivt/}} framework for the 2021 IEEE ICASSP Signal Processing Grand Challenge (SPGC) -- ``COVID-19 Radiomics''. This challenge dataset has been constructed to motivate machine learning practitioners to develop robust and reliable systems to classify patients into COVID-19, CAP and NORMAL diagnosis classes using a heterogeneous set of CT scans. In particular, these CT scans are composed of different slice thicknesses, radiation doses, and noise levels, in addition to featuring patients with various comorbidities and different surgical histories. 

While volumetric CT scans provide a comprehensive illustration of lung abnormalities and their structure, patient level diagnosis from heterogeneous CT volumes faces several challenges as noise and variation between scans can lead to misclassification of individual CT slices. Hence, simplistic score-level/ feature-level \cite{heidarian2020ct, mohammadi2020diagnosis} aggregation performs poorly as there is a tendency for some slices to be misclassified. 
Structures such as 3D-CNNs have also been used to regress volumetric CT inputs directly to the final diagnosis decision \cite{zheng2020deep, zunair2020uniformizing, polat2019classification}. While this allows the model to extract and operate over feature vectors that represent the entire lung of the patient, these models have a very high-dimensional parameter space (tens of millions of trainable parameters) and are prone to over-fitting when trained using datasets with patients (individual samples) in the order of hundreds. 

To alleviate these challenges we propose a novel two-stage framework where features from individual slices are aggregated to a patient level diagnosis via an efficient, light-weight 1D-CNN based model. As novel contributions, (1) our design exploits slice-level features from adjacent slices at different granularities, combining and compressing these discriminative features, prior to classification; (2) has fewer trainable parameters, enabling effective training from a smaller set of volumetric CT scans; (3) our method allows us to seamlessly process examples with a variable number of slices, and even allows the model to learn from incomplete/partial scans; and (4) due to the use of a pre-trained backbone (feature extractor) to extract features from the individual CT slices, the backbone can be swapped or modified. Hence, the proposed two-stage framework is not limited to CT lung classification tasks, but can be easily adapted to any diagnosis task which requires aggregation of heterogeneous information across different samples.

\section{SPGC COVID-19 Radiomics Dataset}
\label{sec:dataset}
The SPGC COVID-19 Radiomics Dataset is one of the largest datasets containing COVID-19, Community Acquired Pneumonia (CAP), and normal cases, and is captured in different medical centers with various imaging settings. The dataset comprises volumetric CT scans of 307 patients (171 COVID-19, 60 CAP, and 76 NORMAL patients). All captured slices in the CT scans are in the Digital Imaging and Communications in Medicine (DICOM) format. The data is acquired using a SIEMENS, SOMATOM Scope scanner with the normal radiation dose and the slice thickness of 2mm. Apart from this patient level labelling, a small subset (i.e 55 COVID-19, and 25 CAP) were analyzed and the individual slices were labeled to indicate evidence of infection. In total 4,993 slices were identified as being indicative of infection. 
From this dataset, 30\% of the data was randomly selected and provided as a validation set. The validation set contains 98 patients (55 COVID, 19 CAP, and 24 NORMAL). The test set consists of three subsets where they consist of 35 COVID, 20 CAP, and 35 NORMAL patients. Test dataset labels are withheld, however, we report the challenge evaluation released by the organisers. 

\section{Methodology}
\label{sec:methods}

\begin{figure*}[htbp]
    \centering
    \includegraphics[width=\textwidth]{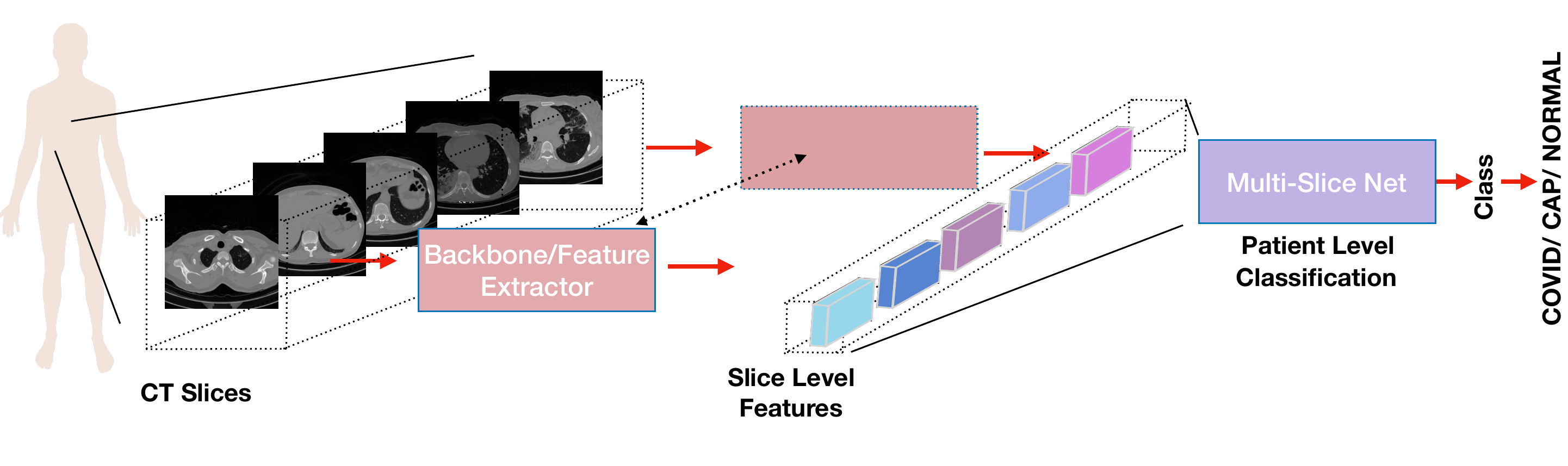}
    \vspace{-12mm}
    \caption{Overall Framework: Individual volumetric chest CT slices are passed through a backbone network for slice level feature extraction. The resultant features are aggregated by the proposed Multi-Slice Network to obtain a patient level diagnosis.}
    \label{fig:overall}
\end{figure*}

We propose a deep network approach, Multi-slice Net, which performs the lung infection classification from the volumetric chest CT scans. The proposed framework is shown in Fig. \ref{fig:overall}, and is composed of a backbone for slice level feature extraction and a network to aggregate these features from a patient to a single score (Multi-Slice Network). 

\subsection{Feature Extractor/ Backbone}
\label{sec:backbone} 
One of the key motivations of the proposed approach is to minimise pre-processing. Hence, aside from converting individual DICOM files to JPG format, no pre-processing steps are performed. In contrast to existing state-of-the-art approaches \cite{heidarian2020ct, zheng2020deep} which perform lung detection and segmentation during pre-processing, the proposed framework applies the feature extractor directly to the JPG slice images.

Extracting features from CT slices that capture discr-iminative infection-related information is crucial for infection classification. We utilise the squeeze-and-excitation ResNet50 (SE-ResNet50) model \cite{hu2019squeeze}, pre-trained on the ImageNet dataset \cite{russakovsky2015imagenet}.The SE-ResNet50 extends the original ResNet50 architecture with the aid of squeeze and excitation operations. In particular, the squeeze operation extracts global information from each of the channels of the input while the excitation act as a bottleneck, adaptively recalibrating the importance of each channel. We fine-tune the SE-ResNet50 model, though the first 6 layers are frozen. For fine-tuning, the subset of patients with slice level annotations are used. This subset contains 55 COVID, and 25 CAP patients. We also randomly selected slices from 15 NORMAL patients for the fine-tuning data. The constructed dataset contains of 2482 COVID, 742 CAP, and 1820 NORMAL slices for training and 1333 COVID, 436 CAP and 840 NORMAL slices for validation. For the compatibility with the pre-trained backbone network, the input CT slices of shape $512\times512\times1$ are resized to $224\times224\times1$ and replicated 3 times ($224\times224\times3$), before being fed to the backbone SE-ResNet50. To reduce over-fitting we used data augmentation and added Random Horizontal Flips with 50\% probability, and randomly changed the brightness, contrast and saturation of the input by a factor of upto 0.4. The network is trained using the Adam \cite{kingma2014adam} optimiser with a learning rate of $1e{-5}$ using Categorical Cross-Entropy Loss for 100 epochs. We used class weights to balance the impact of the minority classes. 

After fine-tuning, we use the model with best validation accuracy and extract the features from the penultimate layer of SE-ResNet50, with a feature dimensionality of 2048.

\subsection{Multi-Slice Network}

\begin{figure}[!h]
    \centering
    \includegraphics[width=\linewidth]{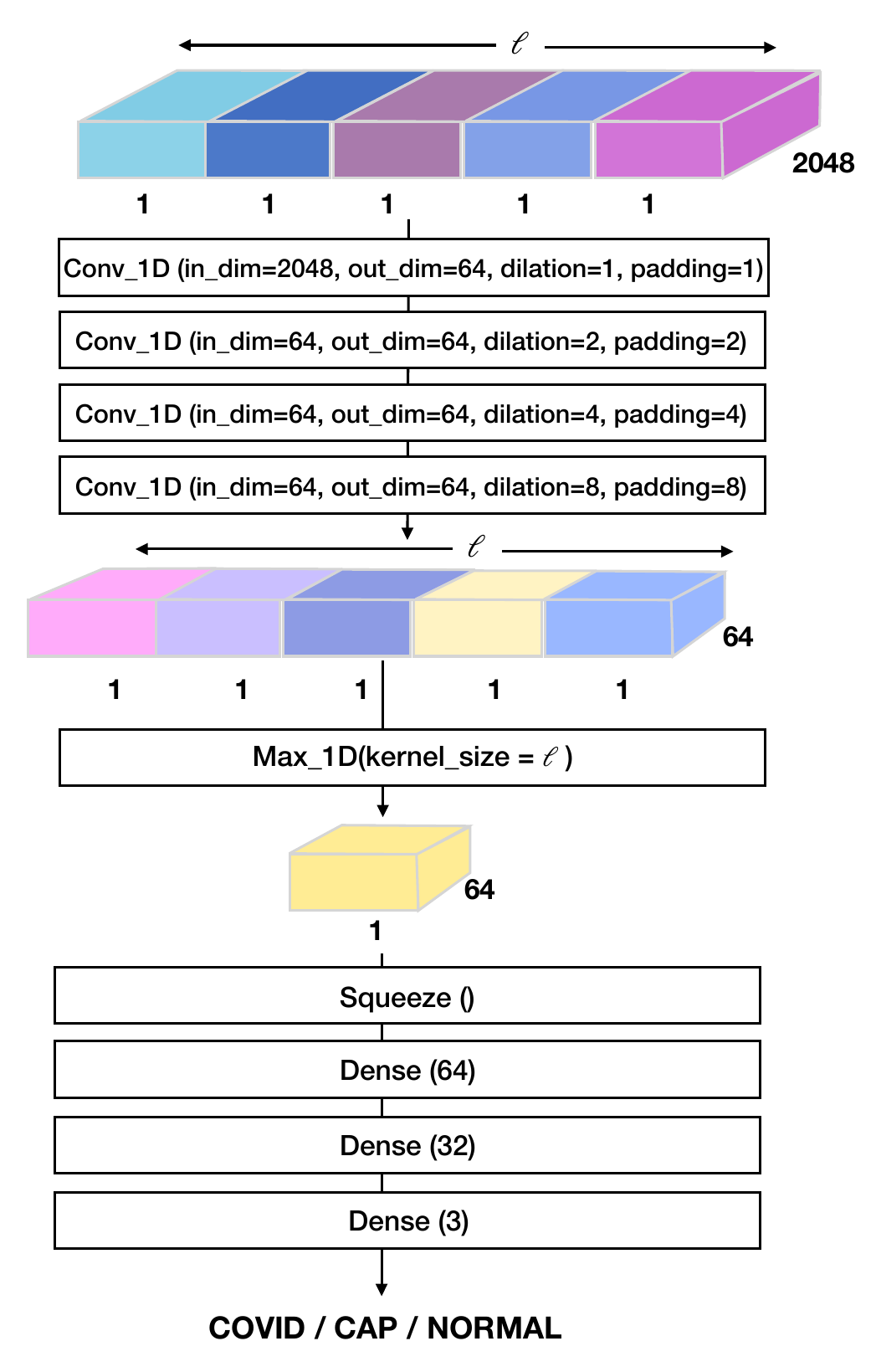}
    \vspace{-10mm}
    \caption{Multi-Slice Network} 
    \label{fig:MSN}
\end{figure}

The features extracted from the backbone then fed to the proposed Multi-Slice Net to obtain a patient-level infection classification. Fig. \ref{fig:MSN} illustrates our approach. Multi-Slice Net iterates through the features extracted from all the CT slices that belong to a particular patient, aggregating them, and generates a single feature descriptor representing the patient. This feature is then fed through a series of dense layers to obtain the final patient diagnosis. 

In the challenge dataset, the CT scans of each patient have a varying number of slices. As such, we designed the network to handle a variable number of slices with the aid of fully convolution network. Let the number of CT slices for a particular subject be $l$, then, the input (I) of the Multi-Slice Net takes the shape ($l$, 2048). Our Multi-Scale Net consists of a temporal convolution layer followed by 4 dilated residual blocks, composed of dilated convolutions. Inspired by \cite{oord2016wavenet, farha2019ms}, we doubled the dilation factor at each layer, and the number of convolutional filters used at each layer is 64. The output of the fourth dilated residual block is then passed through a max-pooling layer with the kernel size of $l$, which encodes the input sequence into a single feature with dimensionality of 64. This feature is then passed through the classification network which is composed of two dense layers with sizes 32 and 3 (number of infection classes) respectively. The use of temporal convolution allows our network to interrogate the slice level features at different granularities, comparing and contrasting features of neighbouring slices. By aggregating these features to a single vector, the most salient features from the patient are passed to the classifier. This network is trained using the Adam \cite{kingma2014adam} optimiser with a learning rate of $1e{-4}$ using Categorical Cross-Entropy Loss for 100 epochs.

\section{Evaluation Results}
\label{sec:evals}

In this section, we first present evaluation results for the fine-tuning process of the feature extractor (Sec. \ref{sec:fintuning_evals}). In Sec. \ref{sec:patient_level_evals} we report patient level diagnosis performance using Multi-Slice Net (MS-Net). 

\subsection{Slice Level Classification Performance (Backbone Networks)}
\label{sec:fintuning_evals}

We evaluate several network architectures to determine an appropriate backbone for feature extraction. When fine-turning these networks, we initialised them with their respective ImageNet weights and fine-tuned them for 100 epochs using the Adam optimiser, a learning rate of $1e{-5}$ and the categorical cross entropy loss. Note that for the fine-tuning process we utilised a subset of the SPGC COVID-19 Radiomics Dataset provided by the organisers which had slice level annotations (see Sec. \ref{sec:backbone} for details). 

\begin{table}[htbp]
\resizebox{\linewidth}{!}{
\begin{tabular}{|c|c|c|c|c|}
\hline
\multirow{2}{*}{Method} & \multicolumn{3}{c|}{Validation Sensitivity}      & \multirow{2}{*}{Validation Accuracy}                        \\ \cline{2-4} 
                                             & COVID           & CAP             & NORMAL          &          \\ \hline
DenseNet \cite{huang2017densely}               & 32.80\%               &  83.66\%              & 71.52\%               &   63.52\%               \\ \hline
ResNet-18 \cite{he2016deep}              & 60.84\%               &  61.39\%              & 90.82\%               &  71.02\%                \\ \hline
SqueezeNet \cite{iandola2016squeezenet}              & 76.72\%          & 71.06\%           & 89.95\%           & 79.15\%  \\ \hline
ResNet-50 \cite{he2016deep}              & 72.08\%                & 78.47\%                & 88.80\%               & 79.92\%                 \\ \hline
SE-ResNet-50  \cite{hu2019squeeze}          & \textbf{79.50\%}               & \textbf{84.58\%}               &  \textbf{96.02\%}             & \textbf{86.63\%}                 \\ \hline
\end{tabular}}
\caption{Slice-level classification accuracy using CT slices from a subset of the SPGC COVID-19 Radiomics dataset. We report class-level sensitivity for COVID, Community Acquired Pneumonia (CAP), and NORMAL classes and overall accuracy (percentage of correct predictions).}
\label{tab:backbone_evals}
\end{table}

Tab. \ref{tab:backbone_evals} provides results for the ResNet-18 \cite{he2016deep}, ResNet-50 \cite{he2016deep}, SqueezeNet \cite{iandola2016squeezenet}, DenseNet \cite{huang2017densely} , and SE-ResNet-50 \cite{hu2019squeeze}  architectures when fine-tuned to obtain a slice level diagnosis. We observe superior performance from the SE-ResNet-50 architecture, despite of the fact that it has been introduced for channel level feature re-calibration on RGB inputs. Despite the need to replicate a single channel CT slice image three times to satisfy the 3-channel requirement of the network, we observe a significant performance increase between ResNet-50 and SE-ResNet-50. We believe this is a result of the removal of redundant/replicated information in channels through the squeeze and excitation blocks of SE-ResNet-50, allowing the classification layers to better focus on informative spatial attributes of the input. 

\subsection{Patient-Level Evaluation}
\label{sec:patient_level_evals}

Evaluation results with respect to the validation set of SPGC COVID-19 Radiomics dataset are provided in Tab. \ref{tab:patient_level_evals}. We report the results of the baseline model provided by the challenge organisers as well as results for MS-Net with different backbones. Our framework outperforms the baseline system, especially when considering the COVID detection sensitivity. We observe similar performance between the ResNet and SE-ResNet backbones, despite the significant performance gap between these methods with respect to slice level evaluations. In Tab. \ref{tab:test_evals} we provide results across testing subsets of the SPGC COVID-19 Radiomics dataset. Despite the lightweight architecture we observe that our framework has achieved competitive results for all classes across all subsets. As the ground truth labels of the test data is not available we cannot compare our performance with existing state-of-the-art models. However, we note that this framework achieved 9th place (from 17 competitive systems) in the SPGC COVID-19 Radiomics challenge. Furthermore, one important characteristic of the proposed method is its consistent performance across the different classes. Despite the heterogeneous test sets, including different slice thicknesses, radiation dose, patient level diff-erences, our lightweight system has been able to achieve consistent performance.   

\begin{table}[!h]
\centering
\resizebox{\linewidth}{!}{
\begin{tabular}{|c|c|c|c|c|}
\hline
\multirow{2}{*}{Method} & \multicolumn{3}{c|}{Validation Sensitivity}      & \multirow{2}{*}{Validation Accuracy}                        \\ \cline{2-4} 
                                             & COVID           & CAP             & NORMAL          &          \\ \hline
Baseline (Provided by Challenge Organisers)  & 42.10 \%        & \textbf{94.5\%}          & 75.00\%         & 79.60\% \\ \hline
MS-Net with ResNet-50 backbone               & \textbf{87.76\%}         & 86.67\%        & 77.27\%         & \textbf{84.88\%} \\ \hline
MS-Net with SE-ResNet-50 backbone            & \textbf{87.76\%}         & 66.67\%         & \textbf{90.91\%}         & \textbf{84.88\%}  \\ \hline
\end{tabular}}
\vspace{-2mm}
\caption{Patient-level evaluations using the validation set of the SPGC COVID-19 Radiomics dataset. We report class-level sensitivity scores for the COVID, Community Acquired Pneumonia (CAP), and NORMAL classes as well as the overall accuracy in terms of the percentage of correctly predicted observations.}
\label{tab:patient_level_evals}
\end{table}

\begin{table}[!h]
\centering
\resizebox{.85\linewidth}{!}{
\begin{tabular}{|c|c|c|c|c|}
\hline
Test subset & COVID & CAP   & NORMAL & Total \\ \hline
Test set 1  & 13/15 & NA    & 14/15  & 27/30 \\ \hline
Test set 2  & 4/10  & 10/10 & 5/10   & 19/30 \\ \hline
Test set 3  & 7/10  & 9/10  & 10/10  & 26/30 \\ \hline
\end{tabular}}
\caption{Patient-level evaluations on different test subsets of the SPGC COVID-19 Radiomics dataset. We report the number of correct identifications against the total ground truth examples for each class. NA refers to Not Applicable as no examples were present in that particular subset.}
\label{tab:test_evals}
\end{table}

Another noteworthy aspects of the proposed approach is the ability to seamlessly switch between different backbone networks. Due to our two-stage approach, the architecture of MS-Net does not require any changes when changing the backbone feature extractor. Moreover, the backbone can be trained in a separate dataset, even without any patient-level data (i.e multiple-slices per patient). As the proposed MS-Net has fewer trainable parameters it can be tuned later with a small scale dataset with fewer patient-level annotations. In addition, we highlight that the MS-Net architecture is not limited to slice level feature aggregation from CT scans. It could be utilised for any aggregation task where features from different spatial or temporal locations need to be aggregated. 

\subsection{Network Complexity}

The majority of the trainable parameters in our framework lie within the backbone feature extractor (SE-ResNet-50), which has 2.5 million trainable parameters (the first six layers are frozen during fine-tuning). MS-Net has only 207,683 trainable parameters due to its careful design. Despite the parameter heavy design of the backbone, the plug and play nature of MS-Net allows the backbone to be pre-trained on a completely different data corpus, and fine-tuned for the task at hand using a smaller dataset. It generates 268 patient level predictions (each of which has a variable number of slices, between 100 and 200, per patient) in 166.9619 seconds. This includes inference for both the backbone network for feature extraction and MS-Net to obtain patient-level predictions. Therefore, on average it takes only 0.6229 seconds to process a CT volume. In future works we will be investigating better backbone architectures to further improve our model's performance, while maintaining it's light weight nature. 

\vspace{-2mm}
\section{Conclusion}
\label{sec:conclusion}
We present a novel light weight framework for COVID-19 diagnosis. Our approach uses a two-stage architecture, composed of a backbone network for feature extraction from individual CT scan slices, and a network to aggregate these slice level features for patient-level diagnosis. Considering the limited data availability of complete patient-level CT volumes, we design a light-weight network to aggregate the slice-level features for patient-level diagnosis. This system is evaluated using the SPGC COVID-19 dataset and achieves competitive results. One prominent attribute of our design is the plug-and-play nature of the aggregation network, which allows the backbone to be trained on a completely different dataset and then tuned on a smaller dataset for the task at hand with patient-level annotations. Future work will include investigation of other backbone designs to further improve model accuracy while maintaining its light weight nature.

\bibliographystyle{IEEEbib}
\bibliography{refs}

\end{document}